\documentclass[%
 reprint,
superscriptaddress,
nofootinbib,
 amsmath,amssymb,
 aps,prl,
]{revtex4-2}

\usepackage{graphicx}
\usepackage{dcolumn}
\usepackage{bm}

\usepackage{color}
\RequirePackage[colorlinks=true
,urlcolor=blue
,anchorcolor=blue
,citecolor=blue
,filecolor=blue
,linkcolor=blue
,menucolor=blue
,linktocpage=true
,pdfproducer=medialab
,pdfa=true
]{hyperref}
\newtheorem{thm}{Theorem}

\begin{document}
\title{Averaged energy conditions for vector fields}
\author{Francisco Jos\'{e} Maldonado Torralba}
\affiliation{Laboratory of Theoretical Physics, Institute of Physics, University of Tartu, W. Ostwaldi 1, 50411 Tartu, Estonia}
%

%
%
\begin{abstract}
In this work we shall obtain sufficient conditions for the appearance of singularities in gravitational theories which propagate an extra vector degree of freedom, based on the known relaxations of the singularity theorems. We study the cases of a general Proca field and a vector theory with stable self-derivative interactions. In this study we show several cases of singularities that usually would be considered as potentially singularity-free, since they violate the usual point-like energy conditions.  
\end{abstract}
%
\maketitle
%
%

\textit{Introduction.}--- 
Every physical theory has a certain range of validity, meaning that it will be able to give a theoretical context to the observations up to a certain energy scale. The theoretical failure of a theory, for example showing divergent quantities, is usually known as a singularity. We can find a very illustrative example in classical electromagnetism, where we observe a divergence of the Coulomb potential at the origin. This is indeed signaling the limited range of the theory.

In gravitational theories, giving a concrete definition of a singularity is more tricky. We can think that a good signal for singular behaviour is when the components of the tensors that describe the curvature of the spacetime diverge. Unfortunately, there are some cases where such a divergence is just a consequence of the chosen coordinates, as it happens in the ``singularity" in $r=2M$ in the Schwarzschild metric. Therefore, we need to resort to another criteria, first proposed by Penrose in \cite{Penrose:1964wq}, which is geodesic incompleteness. Physically this would mean that there are free falling observers that appear or disappear out of nothing, which we can agree it is a singular behaviour. 

The mathematical results which give sufficient conditions for the appearance of singularities in gravitational theories are known as the singularity theorems. The first approach to obtain such a result was made by Raychaudhuri~\cite{Raychaudhuri:1953yv} in 1955, where he introduced his famous equation, proven essential in the later singularity theorems. Ten years later, Penrose formulated the first singularity theorem that does not assume any symmetry \cite{Penrose:1964wq} (see also \cite{Senovilla:2014gza}). The theorem proved that the singularity in $r=0$ of the Schwarzschild metric is also present under non-symmetrical gravitational collapse. Something analogous occurs with the singularity in $t=0$ present in some FLRW metrics, as Hawking showed a year later \cite{Hawking:1966jv}. There it was proven that, under three reasonable conditions, all past directed timelike geodesics have finite length. 

In general, all singularity theorems follow the same structure, made explicit by Senovilla in \cite{Senovilla:1998oua}:
\begin{thm} 
(Pattern singularity ``theorem"). If the spacetime satisfies:\\
1) A condition on the curvature. \\
2) A causality condition. \\
3) An appropriate initial and/or boundary condition. \\
Then there are null or timelike inextensible incomplete geodesics. 
\end{thm}

As it can be seen above, the singularity theorems only depend on geometrical properties of the spacetime. Therefore, they are completely theory independent. Nevertheless, the first condition can be rewritten using the field equations of the considered theory, arriving at the so-called \emph{energy conditions}. Then, the condition written is this form is of course theory dependent.

When studying the possible singular or regular character of gravitational theories beyond General Relativity (GR), it is customary to work directly with the energy conditions. The usual approach is to check if one can break the energy conditions of the theorems, and establish that the theory has the potential to remove the singularities. Such a method presents two very important shortcomings. First, not meeting the conditions of the singularity theorems does not mean that the theory does not exhibit such a pathological behaviour. Secondly, violating the conditions of the modern singularity theorems is actually quite a hard task, since all the conditions of the original theorems have been relaxed. In particular, the curvature or energy conditions can be written in an integral form, so they just need to be met ``in average''.

We will study the case of General Relativity plus an extra vector field, proving several situations where the energy conditions of the original singularity theorems do not hold, but there is still a singularity, since the \emph{average energy conditions} will be met. This result, applied to a common matter source, intends to bring closer the results of the modern singularity theorems to physicists, showing explicitly the strength of the relaxed conditions.

The letter is structured as follows. First, we will briefly review the relaxed version of the Hawking theorem proposed and proved in \cite{Fewster:2010gm}. Then, we will apply such a theorem for the cases of a general Proca field and a vector with self-derivative interactions. Finally, we will give the main conclusions and consequences of the findings.\\

\textit{Relaxed Hawking theorem.}--- 
As we have mentioned, the curvature or energy conditions of the singularity theorems can be relaxed by expressing them in an integral form. Several relaxations have been proposed, but we will focus on the relaxed version of Hawking theorem by Fewster and Galloway \cite{Fewster:2010gm}. In such a reference, they introduce and prove the following singularity theorem,\footnote{In such a reference, a relaxed version of the Penrose theorem is also introduced. Hence, a similar study as the one presented here could be done for the relaxed Penrose theorem.} which we will just outline in the following:
\begin{thm}
\label{thm:Hawking}
Let $(M,g)$ be a globally hyperbolic spacetime of dimension $n\ge 2$, and let $\Sigma$ be a smooth compact spacelike Cauchy surface for M. Assume that along each future complete unit-speed timelike geodesic $\gamma : \left[0,\infty\right)\rightarrow M$, emerging orthogonally from $\Sigma$, there exists $c\geq 0$ such that
\begin{equation}
    \liminf _{T \rightarrow \infty} \int_0^T e^{-2 c t /(n-1)} R_{\mu\nu}v^\mu v^\nu d t>\theta(p)+\frac{c}{2},
\end{equation}
where $R_{\mu\nu}$ is the Levi-Civita Ricci tensor, $v^\mu=\gamma'^\mu(t)$ is the point-like velocity of the geodesic, and $\theta(p)$ is the expansion (see e.g. \cite{Wald:1984rg}) of $\Sigma$ at $p=\gamma(0)$.  
Then M is future timelike geodesically incomplete.
\end{thm}

Such a theorem will allow us to study different vector-tensor modifications of GR and explore the possible appereance of singularities. We shall start by the simplest case, which is GR plus a Proca\footnote{Or Maxwell if the mass of the vector field is zero.} field.\\

\textit{Einstein-Proca theory.}---
The action of a general Einstein-Proca theory in 4 dimensions is given by
\begin{equation}
\label{eq:EPaction}
    S_{\rm{EP}}=\int{\rm{d}}x^4\sqrt{-g}\left(\frac{1}{2\kappa}R+\alpha F_{\mu\nu}F^{\mu\nu}+\beta A_{\mu}A^{\mu}\right),
\end{equation}
where as usual $\kappa=8\pi G$, $A_{\mu}$ is the vector field, $F_{\mu\nu}\equiv\partial_{\mu}A_{\nu}-\partial_{\nu}A_{\mu}$ is the kinetic term, and $\alpha$ and $\beta$ are constants, which we have decided to keep undetermined to also take into account unstable cases. One can always recover the usual Einstein-Proca by setting $\alpha=-\frac{1}{4}$ and $\beta=-\frac{1}{2}m^2$, where $m$ is the mass of the Proca vector field \cite{Heisenberg:2018vsk}.

The Einstein field equations of this theory are 
\begin{eqnarray}
    G_{\mu\nu}&=&\kappa\Bigl[\alpha\bigl(g_{\mu\nu}F_{\rho\sigma}F^{\rho\sigma}-4F_{\mu}\,^{\sigma}F_{\nu\sigma}\bigr) \nonumber \\
    &&+\beta\bigl(g_{\mu\nu}A_{\rho}A^{\rho}-2A_{\mu}A_{\nu}\bigr)\Bigr].
    \label{eq:Einst1}
\end{eqnarray}
As it is customary, in order to obtain the energy condition associated to the curvature condition of Theorem \ref{thm:Hawking}, we shall take the trace of Eq. \eqref{eq:Einst1} and substitute the value of the Ricci scalar back into the equation, and then contract with respect to the unit timelike vector $v^{\mu}$. After some manipulation we arrive at the expression that let us relate the point-like curvature condition and the point-like energy condition
\begin{eqnarray}
    R_{\mu\nu}v^\mu v^\nu&=&-\kappa\Bigl[\alpha\bigl(F_{\rho\sigma}F^{\rho\sigma}+4F_{\mu}\,^{\sigma}F_{\nu\sigma}v^\mu v^\nu\bigr) \nonumber \\
    &&+2\beta\bigl(A_{\mu}A_{\nu}v^\mu v^\nu\bigr)\Bigr],
    \label{eq:CurEn}
\end{eqnarray}
where we have used the fact that $v_{\mu}v^{\mu}=-1$.

At this point, we can study under which values of the parameters and the vector field the condition of the relaxed Hawking theorem can be fulfilled or violated. To obtain the conditions with respect to specific values of the vector field we shall use an orthonormal basis in the local Lorentz frame, as done in \cite{Maeda:2018hqu}. Such a basis is a set of 4 vectors 
\begin{equation}
    e_{a}^\mu=\left(e_{0}^\mu, e_{1}^\mu, e_{2}^\mu, e_{3}^\mu\right),
\end{equation}
which satisfy that
\begin{equation}
    e_{a}^\mu e^{b}_{\mu}=\delta_{a}^{b}.
\end{equation}
The spacetime metric can be constructed as
\begin{equation}
    g_{\mu \nu}= e_\mu^{a} e_\nu^{b}\eta_{ab},
\end{equation}
where $\eta_{ab}=\operatorname{diag}(-1,1,1,1)$ is the metric in the local frame. This kind of basis is also known in the literature as \emph{vilbein}. Any tensor $B_{\mu_1\mu_2\ldots\mu_n}$ can be expressed in components under this basis as
\begin{equation}
    B_{\mu_1\mu_2\ldots\mu_n}=B_{a_1 a_2\ldots a_n}e_{\mu_1}^{a_1} e_{\mu_2}^{a_2}\ldots e_{\mu_n}^{a_n}.
    \label{eq:components}
\end{equation}

In order to calculate expression \eqref{eq:CurEn} in the orthonormal basis, we shall first take into account that for any unit timelike vector $v^{\mu}$ we can perform a Lorentz transformation to set the frame such that $v^{a}=(1,0,0,0)$, without loss of generality. Moreover, we can express the vector field $A_\mu$ using the prescription in Eq. \eqref{eq:components} and the kinetic term as \cite{Maeda:2018hqu}
\begin{equation}
    F_{\mu\nu}=2 \sum_{i=1}^{3} F_{0 i} \,e_\mu^{[0} e_v^{i]}+2 \sum_{i=1}^{3} \sum_{j>i}^{3} F_{ij} e_\mu^{i} e_v^{j}\,,
\end{equation}
where the antisymmetrisation $[\ldots]$ is defined with the customary normalization factor. Also, let us note that along this work only the Latin indices will be the ones acquiring particular values. 

With the previous we can calculate each term in \eqref{eq:CurEn}. For the ones involving the kinetic term, we first obtain
\begin{equation}
     F_{\mu \nu}v^\mu=\sum_{i=1}^{3} F_{0i} \,e_\nu^{i},
\end{equation}
which is clearly a spacelike vector. Hence, we can still use the freedom that we have in the spacelike section of the basis to perform a spacelike rotation such that $F_{\mu \nu}v^\mu$ is in the direction of $e_\nu^{1}$. In such a frame we have that $F_{02}=F_{03}=0$, and consequently 
\begin{equation}
     F_{\mu \nu}v^\mu=F_{01} \,e_\nu^{1}.
\end{equation}
Moreover, in the same frame we have
\begin{equation}
    F_{\mu \nu} F^{\mu \nu}=-2 (F_{01})^2+2 \sum_{i=1}^{3} \sum_{j>i}^{3} (F_{ij})^2.
\end{equation}

With all the previous considerations we can express Eq. \eqref{eq:CurEn} as 
\begin{eqnarray}
    R_{\mu\nu}v^\mu v^\nu&=&-2\kappa\biggl\{\alpha\Bigl[(F_{01})^2+ \sum_{i=1}^{3} \sum_{j>i}^{3} (F_{ij})^2\Bigr] \nonumber \\
    &&\qquad+\beta(A_{0})^2\biggl\},
    \label{eq:EnCondProca}
\end{eqnarray}

First of all, it is clear that for the standard Proca values of the parameters the point-like energy condition holds, and hence the averaged one. The only possibility of breaking the energy condition is that either the vector is a ghost, $\alpha>0$, or a tachyon, $\beta>0$ and $\alpha<0$, both of them being unstable cases. In the following we shall prove that in both cases, which would be usually rendered as potentially singularity-free, the averaged energy condition can be met and hence there is a singularity. 

Before studying the specific cases let us consider a general situation where the point-like convergence condition is expressed as 
\begin{equation}
    R_{\mu\nu}v^\mu v^\nu=G+H,
    \label{eq:general}
\end{equation}
where $G$ and $H$ are functions meeting that $G\ge 0$ and $H<0$, which means that the point-like energy condition can be violated. Following the arguments of \cite{Fewster:2010gm} for the scalar field case, let us consider the Cauchy hypersurface $\Sigma$ of the singularity theorem \ref{thm:Hawking}, and allow $H$ to have an exponential increase but bounded by 
\begin{equation}
    \left|H\right|\leq h\cdot{\rm e}^{\lambda d(p)}
    \label{eq:genbound}
\end{equation}
for any point $p$ in the causal future of $\Sigma$. Here $h$ and $\lambda$ are positive constants, and $d(p)$ is the (Lorentzian) distance from $p$ to $\Sigma$. It is important that the distance to the hypersurface is present since, due to the properties of Cauchy hypersurfaces, there is always an unit-speed geodesic $\bar{\gamma}(t)$ that gives the distance to the hypersurface from any of its points as $d(\bar{\gamma}(t))=t$. 

Now, considering such a geodesic we can calculate the following bound from \eqref{eq:general} and \eqref{eq:genbound}
\begin{eqnarray}
    &&\int_0^T e^{-\frac{2}{3} c t} R_{\mu\nu}v^\mu v^\nu d t\geq \int_0^T e^{-\frac{2}{3} c t} H d t \nonumber \\
    &&\quad\geq -h\int_0^T {\rm e}^{\left(\lambda-\frac{2}{3} c\right) t} d t\geq -h\int_0^\infty {\rm e}^{\left(\lambda-\frac{2}{3} c\right) t} d t \nonumber \\
    &&\quad=-h\left(\frac{2}{3} c-\lambda\right)^{-1},    
    \label{eq:bound}
\end{eqnarray}
which holds for any $T>0$ and $\lambda<\frac{2}{3} c$.
Let us note that the previous bound will apply for every unit-speed geodesic since $\bar{\gamma}(t)$ maximizes the distance from the points to the hypersurface. 

Using the obtained inequality \eqref{eq:bound}, we can realize that we can find cases for which there are singularities even if $R_{\mu\nu}v^\mu v^\nu$ is negative and its absolute value is up to exponentially increasing. In order to see this, in the energy condition of Theorem \ref{thm:Hawking} and in Eq. \eqref{eq:bound} we shall consider $c=\frac{3}{4}\lambda+\frac{3}{2}\sqrt{\frac{\lambda^{2}}{4}+\frac{4}{3}h}$. Then, comparing both we can clearly see that the averaged energy condition is fulfilled if $\theta<-c$.   

With the general case explained, we can go back to \eqref{eq:EnCondProca} and study the ghost and tachyon in a very straightforward way. On the one hand, it is clear that for the tachyon field, $\beta>0$ and $\alpha<0$, we can compare with \eqref{eq:general} and identify the kinetic part of \eqref{eq:EnCondProca} with $G$ and the second term with $H$. This means that if $A_0$ fulfills a bound like the one in \eqref{eq:genbound} there will still be a singularity, even if the point-like energy condition does not hold. On the other hand, for the ghost field, $\alpha>0$, we can distinguish two cases depending on the sign of $\beta$. If $\beta<0$, then we can identify the $A_0$ term with $G$ and the kinetic part with $H$. Therefore if the kinetic part follows the bound \eqref{eq:genbound} the spacetime will be singular. If $\beta>0$, we have that all the terms need to meet the bound in order to still have a singularity.\\

\textit{Self-derivative interactions.}---
Now we will explore the case where we allow stable self-derivative interactions in the vector-tensor theory. The action considered is the following
\begin{equation}
\label{eq:SDaction}
    S_{\rm{SDI}}=S_{\rm{EP}}+\int{\rm{d}}x^4\sqrt{-g}\left(C_1 A_{\mu}A^{\mu}\nabla_{\nu}A^{\nu}\right),
\end{equation}
where $C_1$ is a constant and the new term is the only stable self-derivative interaction for a vector field, up to boundary terms \cite{BeltranJimenez:2016rff}. For convenience, we shall introduce the symmetric derivative $S_{\mu\nu}\equiv\nabla_{\mu}A_{\nu}+\nabla_{\nu}A_{\mu}$, and consequently the divergence would be related to its trace as $S\equiv S_{\mu}^{\mu}=2\nabla_{\mu}A^{\mu}$. The field equations of this theory are given by
\begin{equation}
    G_{\mu\nu}=T^{\rm P}_{\mu\nu}-\kappa C_1\biggl[A_{\mu}A_{\nu}S+A_{\rho}A^{\rho}\Bigl(S_{\mu\nu}-\frac{1}{2}g_{\mu\nu}S\Bigr)\biggr],
    \label{eq:Einst2}
\end{equation}
where $T^{\rm P}_{\mu\nu}$ refers to the right-hand side of Eq. \eqref{eq:Einst1}. Again, after some manipulation on the previous equation we arrive at the expression that relates the point-like curvature and energy conditions
\begin{eqnarray}
    R_{\mu\nu}v^\mu v^\nu&=&{\rm Proca}-\kappa C_1\biggl[S A_{\mu}A_{\nu}v^{\mu}v^{\nu}
    \nonumber \\
    &&+A_{\rho}A^{\rho}\Bigl(S_{\mu\nu}v^{\mu}v^{\nu}+\frac{1}{2}S\Bigr)\biggr],
    \label{eq:CurEn2}
\end{eqnarray}
where ``Proca'' refers to the contribution of the Proca part, which is the right-hand side of Eq. \eqref{eq:CurEn}. In order to obtain the specific values of the energy condition, we shall work in the same orthonormal basis as in the Proca case. In such basis, and with the choices we have made, we have that
\begin{equation}
    S_{\mu\nu}v^{\mu}v^{\nu}=S_{00},
\end{equation}
and the trace $S$ will be left without expanding it for convenience. Hence, we expand Eq. \eqref{eq:CurEn2} in term of components as follows
\begin{eqnarray}
    R_{\mu\nu}v^\mu v^\nu&=&-2\kappa\biggl\{\alpha\Bigl[(F_{01})^2+ \sum_{i=1}^{3} \sum_{j>i}^{3} (F_{ij})^2\Bigr] \nonumber \\
    &&+\Bigl[\beta+\frac{\kappa C_{1}}{4}\left(S-2S_{00}\right)\Bigl](A_{0})^2
    \nonumber
    \\
    &&+\frac{\kappa C_{1}}{4}\left(S+2S_{00}\right)\sum_{i=1}^{3}(A_{i})^{2}\biggl\}.
    \label{eq:EnCondSelf}
\end{eqnarray}
It is clear to see that even imposing that the vector field is not a ghost or a tachyon, \emph{i.e.} $\alpha,\beta<0$, we have different possibilities of breaking the point-like energy condition. We shall outline them in the following and compare them with the general case \eqref{eq:general}.
\begin{itemize}
    \item $S-2S_{00}>-\frac{4\beta}{\kappa C_{1}}$ and $S<-2S_{00}$: Then we can identify the term involving $A_0$ with $H$ and the rest with $G$. Hence, if $A_0$ is bounded in the way of \eqref{eq:genbound} there will be a singularity.

    \item $S-2S_{00}<-\frac{4\beta}{\kappa C_{1}}$ and $S>-2S_{00}$: Here we can identify the term involving the spatial components of the vector as $H$ and the rest as $G$. Consequently, if the spatial part is bounded like \eqref{eq:genbound} a singularity will be present.

    \item $S-2S_{00}<-\frac{4\beta}{\kappa C_{1}}$ and $S<-2S_{00}$: We can identify the kinetic part with $G$ and the rest with $H$. Hence, if all the components of the vector follow the bound \eqref{eq:genbound}, there will be a singularity. 
\end{itemize}

In conclusion, by studying the averaged energy conditions in two vector theories, we have proven that there are several singular cases which will be considered potentially singularity-free with respect to the original singularity theorems. This signals the difficulty of actually removing the singularities even when modifying the gravitational theory.\\

\begin{acknowledgments}
The author would like to thank Alejandro Jiménez Cano for useful discussions and feedback. The author is supported by the ``Fundaci\'on Ram\'on Areces''. This research was also supported by the European Regional Development Fund through the Center of Excellence TK133 “The Dark Side of the Universe”. 
\end{acknowledgments}

\bibliographystyle{apsrev}
\bibliography{references}

\end{document}